# Causal Wave Mechanics and the Advent of Complexity.
# I. Dynamic multivaluedness

A.P. K<small>IRILYUK</small>*

Institute of Metal Physics, Kiev, Ukraine 252142

ABSTRACT. Two major deviations from causality in the existing formulations of quantum mechanics, related respectively to quantum chaos and indeterminate wave reduction, are eliminated within the new, universal concept of dynamic complexity. The analysis involves a new paradigm for description of a system with interaction, the principle of dynamic multivaluedness (redundance), and the ensuing concept of the fundamental dynamic uncertainty. It is shown that both the wave reduction and truly unpredictable (chaotic) behaviour in quantum systems can be completely and causally understood as a higher sublevel of the same dynamic complexity that provides the causally complete picture of the unified wave-particle duality and relativity at its lowest level (quant-ph/9902015,16). The presentation is divided into five parts. The first three parts deal with intrinsic randomness in Hamiltonian (isolated) quantum systems as the basic case of dynamical chaos. In the last two parts a causal solution to the problem of quantum indeterminacy and wave reduction is proposed. Part I introduces the method of effective dynamical functions as a generalisation of the optical potential formalism. The method provides a legal transformation of the Schrödinger equation revealing the hidden multivaluedness of interaction process, i. e. its self-consistent, dynamical splitting into many equally real, but mutually incompatible branches, called 'realisations'. Each realisation incorporates the usual "complete" set of eigenfunctions and eigenvalues for the entire problem. The method is presented in detail for the Hamiltonian system with periodic (not small) perturbation, in both its time-independent and time-dependent versions.

NOTE ON NUMERATION OF ITEMS. We use the unified system of consecutive numbers for formulas, sections, and figures (but *not* for literature references) throughout the full work, Parts I-V. If a reference to an item is made outside its "home" Part of the work, the Roman number of this home part is added to the consecutive number: 'section 9.2' and 'section 9.2.IV' refer to the same, uniquely defined section, but in the second case we know in addition that it can be found in Part IV of the work.

---

*Address for correspondence: Post Box 115, Kiev - 30, Ukraine 252030.
 E-mail address: kiril@metfiz.freenet.kiev.ua

# 1. Introduction

The recent emergence and development of the dynamical chaos concept have engendered profound changes in our understanding of both fundamental and practical aspects of the dynamical system behaviour in many different fields of physics (e. g. [1]). In particular, the behaviour of simple non-dissipative mechanical systems with few degrees of freedom presents an elementary case well suited for the study of the fundamental origins of dynamical randomness, with possible further extension to more complex situations. The description of chaos in such elementary dynamical systems within the formalism of classical mechanics has seemed to be rather successful and self-consistent [1-4]. At the same time its proposed quantum-mechanical versions, despite a large amount of the efforts made, have failed in creating a similar prosperous situation, even though a number of important particular results has been obtained [4-7]. The problem of the very existence of the truly unpredictable behaviour of deterministic quantum systems remains unsettled [8]. The fundamental difficulty stems from the unavoidable wave involvement in quantum postulates: waves do not easily show global instabilities necessary for the development of chaotic regimes or, in other terms, waves lead to discreteness (one may physically realise minimum a half-wave and not, say, one fifth of it), and the discreteness is incompatible with the existing notion of instability appealing to infinitesimal values and apparently indispensable for the known definitions of chaos. Solutions to this basic problem, proposed or implied, vary over a wide but already shrinking range.

The most popular point of view involves the reduction of quantum chaos to some kind of very intricate but basically regular behaviour related eventually to the peculiarities of involved mathematical objects like zeta-functions [4,9] or random matrices [10]. Whatever the variations of such approach, it suffers from the evident conflict with the rather attractive idea, strongly supported by the existence of different types of classical chaos, that the world's complexity should be greater than zero: the latter would be impossible without the irreducible 'true' unpredictability and randomness in quantum mechanics as the most general description of the world. If it is not the case, i. e. if, for example, algorithmic complexity of quantum dynamics turns invariably to zero, then one possibility is that the existing quantum mechanics is no more general enough to provide a non-contradictory description for the world of chance [11].

The practical absence, however, of such properly chaotic scheme of quantum mechanics leaves the way open for another possibility: the conventional quantum mechanics itself is valid, for classically chaotic quantum systems in the same sense that for the classically regular ones, while the chaoticity of its classical limit "appears" due to the particularities of the semiclassical transition. The emerging "quantum chaology" [12] finally also proposes rather a possibility than a solution: it is *suggested* why quantum and classical descriptions *need not* provide similar concepts of chaos, but it remains unclear why and *how exactly* the quantum-mechanical reality with complexity zero (absolute dynamical predictability and reversibility) passes to its truly random limiting case with non-zero complexity at the other end of the semiclassical transition. The latter should therefore 'produce' complexity 'from nothing' by a still unknown mechanism.

This gives us also a hint to the third and logically the last remaining possibility: the true solution is hidden neither within the semiclassical transition, nor in a reinterpretation of quantum mechanics, but rather in our understanding of chaos and complexity in general, even in classical mechanics. It may imply that in reality the 'pure' randomness and the corresponding complexity do not exist at all, and the apparent world of chance is just an illusory product of the ever wavering spirit... This choice, rather exotic as it may seem, can be provided, nonetheless, with a precise rational basis: it is stated simply that as the quantum-mechanical "diffusion localisation time" [13], after which a system starts deviating from apparently chaotic behaviour, becomes practically infinitely large for classical objects, one can never have a chance to achieve this regime of chaos suppression in the classical world, contrary to small quantum objects (this opinion was communicated to the author by Dr. S. Weigert from Basel University). The world is thus not complex, within this possibility, neither quantum-mechanically, nor classically, even though it is permitted to be very intricate. It represents, in fact, a huge generator of random numbers, basically deterministic but very perfect in its simulation of randomness (in other words, the world's dynamics is a periodic one, but the period is extremely large). In this way the problem of the origin of quantum chaos is effectively reduced to the alternative between the true complexity and a simple intricacy of being.

Apart from this logical choice among the three basic possibilities, one may mention the hypotheses about the quantum measurement involvement in chaos (see e. g. the special NATO Workshop Proceedings [14]) or the "fundamentally irreducible" representation by the density matrix [15]. It is not difficult to see that these approaches can also be reduced to one of the above possibilities touching the roots of the quantum paradigm. At their present state, however, they seem to be either too abstract, or too special to form a universal basis for the first-principle understanding of quantum chaos in real physical systems.[*)]

In the absence of clearness in the fundamental aspects, there is the general tendency to study particular features, or "signatures", of chaos in the *regular* quantum dynamics of the classically chaotic systems [6,7] leaving aside the related logical puzzles. However, it becomes more and more evident that one cannot hope to be really successful even in such particular research without the

---

[*)] We do not discuss here the appearing multiple attempts to relate deterministic randomness in quantum systems to a specific behaviour of the *dissipative* quantum systems inevitably involving an external source of irreversibility, whether it is stochasticity of the environment or the equivalent "coarse-graining" of a problem. We are looking for an intrinsic causal source of dynamic unpredictability which would remain in the zero-noise limit, even though the addition of noise may play an important practical role in complex system behaviour (see sections 2.3.II and 5.III below). This approach involves eventually considerable extension of the existing interpretation of *any* type of chaotic behaviour (see sections 6.III and 10.V of this work), including the "divergent-trajectories" paradigm of classical deterministic chaos that suffers, in fact, from the same incompleteness as the "dissipative chaos" concept.

general self-consistent picture including the basic issues (not to mention the temptations of pure curiosity). This situation predetermines the importance of the search for a general self-consistent quantum (and eventually any other) chaos description providing the fundamental origin of randomness in deterministic systems. In this essay we present such an approach starting from the application of the unreduced version of the well-known optical potential method (see e. g. [16]) to the analysis of quantum chaos in Hamiltonian systems with periodic perturbation. This approach appeared originally as a part of quantum-mechanical description of charged particle scattering in crystals [17] revealing its chaotic behaviour (see, in particular, section 2.5 of the cited article). The results obtained are considerably developed and generalised in the present paper.

Using a generic example of arbitrary periodically perturbed Hamiltonian system, we show that our method naturally leads to the new concept of the fundamental origin of chaos in dynamical systems (section 2). In particular, it permits one to overcome the 'pathological regularity' of quantum mechanics and to perform the ordinary semiclassical transition also for chaotic systems, in agreement with the correspondence principle (section 3.II). These results show that our approach is presented in the form ready for its application to practical study of Hamiltonian quantum chaos in various real physical systems including both basic aspects and the particular analysis of the measured quantities.

An important topic concerning fractals involvement with chaos is considered in section 4.III, where we show, within the same approach, that fractals naturally appear as solutions of the modified Schrödinger equation for a chaotic system that can be obtained analytically and not only by computer simulations (the result refers, in fact, to any dynamic equation presented in the proper form). In fact, we demonstrate that any solution describing a system in a chaotic regime (and this is the absolute majority of all situations) has a fractal character, and the properties of this fractal can be determined by the outlined procedure.

It is important that in order to obtain quantum dynamics with non-zero complexity and the conventional semiclassical transition, one does not need to reconsider the foundations of quantum mechanics as such but rather to use another, more general, form of the same formalism. This can help to moderate the painful choice described above (we discuss these issues in more detail in section 5.III). Moreover, in parts IV, V of this work we show that it is the main unsolved problems of the foundations of quantum mechanics, known as quantum indeterminacy and wave reduction, that can be given transparent causal solutions by application of the same method to the process of quantum measurement.

Finally, the purpose of this work is to demonstrate the universal character of the results obtained permitting one to extend the same concept of complex behaviour to other Hamiltonian and non-Hamiltonian, classical, and eventually distributed nonlinear systems. The ensuing universal notions of the fundamental dynamic uncertainty, randomness, probability, complexity, (non)integrability, and general solution are introduced in section 6.III.

# 2. Formulation of the method
## 2.1. Effective dynamical functions

Consider a conservative dynamical system with the Hamiltonian

$$H = h + V,$$

where $h$ is the free motion Hamiltonian, and $V$ describes the elastic interaction. Within the total Hamiltonian $H$ we separate the integrable part, $H_0$, corresponding to the regular dynamics, and the perturbation, $H_p$, (generally, not small) inducing chaotic behaviour of the whole system which we want to describe:

$$H = H_0 + H_p, \quad H_0 \equiv h_0 + V_0, \quad H_p \equiv h_p + V_p \quad .$$

We introduce then a particular representation, for definitness chosen in the form of coordinate representation, $H \equiv H(\mathbf{r})$, and divide the vector of independent variables into two parts, $\mathbf{r} = \{\mathbf{r}_\sigma, \mathbf{r}_\pi\}$, so that, in accord with the integrability of $H_0$, $H_0 = H_0(\mathbf{r}_\sigma)$ and $H_p = H_p(\mathbf{r}_\sigma, \mathbf{r}_\pi)$. If chaos is induced by the addition of extra degrees of freedom (dimensions), the variables $\mathbf{r}_\pi$ may correspond to these degrees, while the motion limited to the degrees $\mathbf{r}_\sigma$ is considered to be regular. In cases where such subdivision is not naturally given by the conditions of a problem, it can always be made using the well-known regularity of one-dimensional problems: in the simplest version a one-dimensional component of $\mathbf{r}$ is chosen as $\mathbf{r}_\sigma$, $\mathbf{r}_\sigma = x$, $\mathbf{r} = \{x, \mathbf{r}_\pi\}$. The subdivision of the Hamiltonian is then performed with the help of the Fourier analysis or other suitable expansion. As we shall see later, the case of time-dependent perturbation can also be considered within the same formalism, and then $\mathbf{r}_\pi$ corresponds to the time variable, $t$. If the degrees of freedom remain unchanged and chaos is due to a symmetry-breaking perturbation, one can still use the above formal method of division. To obtain physically more meaningful result, one may imply, where possible, the quantum analogues of the action-angle variables ($\mathbf{r}_\sigma \to \mathbf{I}$, $\mathbf{r}_\pi \to \theta$), or those determined by the symmetry (e. g. $\mathbf{r}_\sigma = r$), or other suitable choice.[*]

While studying purely dynamical origins of stochasticity, it is natural to assume that $V_p$ is a periodic function of $\mathbf{r}_\pi$. It is also one of the most interesting particular cases (e. g. particle scattering in regular structures [17,18], atom excitation by electromagnetic radiation [19], the kicked rotor model [20], etc.), and it is well suited for the demonstration of our method. The dependences of $V_0$ and $V_p$ on $\mathbf{r}_\sigma$ may be periodic or not.

---

[*] We consider it to be always possible. Although our description encompasses, in principle, this latter type of chaos, the results below are specified rather for the former one, with some generalising notes where possible; the detailed investigation of the second type of chaos is left for next publications.

Consider now the Schrödinger equation for our system,

$$[h_0(\mathbf{r}_\sigma) + h_p(\mathbf{r}_\pi) + V_0(\mathbf{r}_\sigma) + V_p(\mathbf{r}_\sigma,\mathbf{r}_\pi)]\Psi(\mathbf{r}_\sigma,\mathbf{r}_\pi) = E\Psi(\mathbf{r}_\sigma,\mathbf{r}_\pi) \,, \qquad (1)$$

where $E$ is the total energy. The involvement of periodicity inspires the idea about the Fourier transformation over $\mathbf{r}_\pi$[*] which is done in the usual way and leads to the system of equations for the component functions $\psi_{\mathbf{g}_\pi}(\mathbf{r}_\sigma)$:

$$[h_0(\mathbf{r}_\sigma) + V_0(\mathbf{r}_\sigma)]\psi_0(\mathbf{r}_\sigma) = \varepsilon_\sigma \psi_0(\mathbf{r}_\sigma) - \sum_{\mathbf{g}_\pi} V_{-\mathbf{g}_\pi}(\mathbf{r}_\sigma)\psi_{\mathbf{g}_\pi}(\mathbf{r}_\sigma) \,, \qquad (2a)$$

$$h_0(\mathbf{r}_\sigma)\psi_{\mathbf{g}_\pi}(\mathbf{r}_\sigma) + \sum_{\mathbf{g}_\pi'} V_{\mathbf{g}_\pi-\mathbf{g}_\pi'}(\mathbf{r}_\sigma)\psi_{\mathbf{g}_\pi'}(\mathbf{r}_\sigma) = \varepsilon_{\sigma\mathbf{g}_\pi}\psi_{\mathbf{g}_\pi}(\mathbf{r}_\sigma) - V_{\mathbf{g}_\pi}(\mathbf{r}_\sigma)\psi_0(\mathbf{r}_\sigma) \,, \qquad (2b)$$

where $\mathbf{g}_\pi, \mathbf{g}_\pi' \neq 0$ are the dual "reciprocal lattice" vectors with respect to the "direct lattice" of vectors $\mathbf{r}_\pi$,

$$\varepsilon_{\sigma\mathbf{g}_\pi} \equiv E - h^2(\mathbf{K}_\pi + \mathbf{g}_\pi)^2/2m \,, \qquad \varepsilon_\sigma \equiv E - h^2 K_\pi^2/2m \,, \qquad (3)$$

$$\Psi(\mathbf{r}) = \exp(i\mathbf{K}_\pi\mathbf{r}_\pi)\left[\psi_0(\mathbf{r}_\sigma) + \sum_{\mathbf{g}_\pi}\psi_{\mathbf{g}_\pi}(\mathbf{r}_\sigma)\exp(i\mathbf{g}_\pi\mathbf{r}_\pi)\right] \,,$$

$$V(\mathbf{r}) = V_0(\mathbf{r}_\sigma) + \sum_{\mathbf{g}_\pi} V_{\mathbf{g}_\pi}(\mathbf{r}_\sigma)\exp(i\mathbf{g}_\pi\mathbf{r}_\pi) \,,$$

and the wave vector $\mathbf{K}_\pi$ corresponds to the standard Bloch representation of the wave function for periodic potentials (see e. g. [16]).

Now one can use this decomposition to study the influence of "chaos bringing" perturbation $V_p(\mathbf{r}_\sigma,\mathbf{r}_\pi)$ providing the terms with $\mathbf{g}_\pi \neq 0$. We start by applying the simple method of substitution and first express $\psi_{\mathbf{g}_\pi}(\mathbf{r}_\sigma)$ through $\psi_0(\mathbf{r}_\sigma)$ from eq. (2b) with the help of the Green function for its homogeneous part with respect to $\psi_{\mathbf{g}_\pi}(\mathbf{r}_\sigma)$,

$$h_0(\mathbf{r}_\sigma)\psi_{\mathbf{g}_\pi}(\mathbf{r}_\sigma) + \sum_{\mathbf{g}_\pi'} V_{\mathbf{g}_\pi-\mathbf{g}_\pi'}(\mathbf{r}_\sigma)\psi_{\mathbf{g}_\pi'}(\mathbf{r}_\sigma) = \varepsilon_{\sigma\mathbf{g}_\pi}\psi_{\mathbf{g}_\pi}(\mathbf{r}_\sigma) \,. \qquad (4a)$$

The Green function is given by the well-known expression:

$$G_{\mathbf{g}_\pi}(\mathbf{r}_\sigma,\mathbf{r}_\sigma') = \sum_n \frac{\psi_{\mathbf{g}_\pi n}^0(\mathbf{r}_\sigma)\psi_{\mathbf{g}_\pi n}^{0*}(\mathbf{r}_\sigma')}{\varepsilon_{\mathbf{g}_\pi n}^0 - \varepsilon_{\sigma\mathbf{g}_\pi}} \,,$$

---

[*] In the general case it will be an expansion in terms of the other complete system of functions appropriate to a problem.

where $\{\psi^0_{\mathbf{g}_\pi n}(\mathbf{r}_\sigma)\}$ and $\{\varepsilon^0_{\mathbf{g}_\pi n}\}$ are the sets of eigenfunctions and eigenvalues, respectively, for the auxiliary system of equations (4). The solution of the system (2b) can be expressed as

$$\psi_{\mathbf{g}_\pi}(\mathbf{r}_\sigma) = - \int_{s_\sigma} d\mathbf{r}_\sigma' \, G_{\mathbf{g}_\pi}(\mathbf{r}_\sigma,\mathbf{r}_\sigma') V_{\mathbf{g}_\pi}(\mathbf{r}_\sigma') \psi_0(\mathbf{r}_\sigma') \;,$$

where the domain of integration $s_\sigma$ coincides with the "unit cell" for $V(\mathbf{r})$ periodic in $\mathbf{r}_\sigma$ or with the whole domain of definition on $\mathbf{r}_\sigma$ for a non-periodic potential.

Now we substitute the obtained expression for $\psi_{\mathbf{g}_\pi}(\mathbf{r}_\sigma)$ in the right-hand side of eq. (2a) and come to the conclusion that the problem is reduced to the *modified Schrödinger equation* for $\psi_0(\mathbf{r}_\sigma)$:

$$[h_0(\mathbf{r}_\sigma) + V_{\text{eff}}(\mathbf{r}_\sigma)]\psi_0(\mathbf{r}_\sigma) = \varepsilon_\sigma \psi_0(\mathbf{r}_\sigma) \;, \tag{5}$$

where the ordinary potential $V_0(\mathbf{r}_\sigma)$ is replaced by the *effective potential* (EP) $V_{\text{eff}}(\mathbf{r}_\sigma)$, also known as coherent, or optical, potential [16]. It is obtained as a sum,

$$V_{\text{eff}}(\mathbf{r}_\sigma) = V_0(\mathbf{r}_\sigma) + \vartheta(\mathbf{r}_\sigma) \;, \tag{6a}$$

where $\vartheta(\mathbf{r}_\sigma)$ is the nonlocal part of EP expressed by the integral operator:

$$\vartheta(\mathbf{r}_\sigma)f(\mathbf{r}_\sigma) \equiv \int_{s_\sigma} d\mathbf{r}_\sigma' \, V(\mathbf{r}_\sigma,\mathbf{r}_\sigma')f(\mathbf{r}_\sigma') \;. \tag{6b}$$

The integral kernel $V(\mathbf{r}_\sigma,\mathbf{r}_\sigma')$ can be presented in the form

$$V(\mathbf{r}_\sigma,\mathbf{r}_\sigma') = \sum_{\mathbf{g}_\pi,n} \frac{V_{-\mathbf{g}_\pi}(\mathbf{r}_\sigma) V_{\mathbf{g}_\pi}(\mathbf{r}_\sigma') \psi^0_{\mathbf{g}_\pi n}(\mathbf{r}_\sigma) \psi^{0*}_{\mathbf{g}_\pi n}(\mathbf{r}_\sigma')}{\varepsilon_\sigma - \varepsilon^0_{\mathbf{g}_\pi n} - \varepsilon_{\pi g_\pi} - 2\cos\alpha_{\mathbf{g}_\pi}\sqrt{(E-\varepsilon_\sigma)\varepsilon_{\pi g_\pi}}} \;, \tag{6c}$$

where

$$\varepsilon_{\pi g_\pi} \equiv \hbar^2 g_\pi^2 / 2m \;,$$

and $\alpha_{\mathbf{g}_\pi}$ is the angle between the vectors $\mathbf{g}_\pi$ and $\mathbf{K}_\pi$ (for one-dimensional $\mathbf{r}_\pi$, $\alpha_{\mathbf{g}_\pi}$ takes only two values, $\alpha_{\mathbf{g}_\pi} = 0,\pi$).

This expression for EP contains, in particular, the unknown solutions of the auxiliary system of equations (4). Their properties are investigated in Appendix (see also [17]), where it is shown that qualitatively they are similar to those of solutions of the Schrödinger equation with the unperturbed potential $V_0(\mathbf{r}_\sigma)$. More detailed quantitative characteristics of these solutions and their properties can be analysed using suitable approximations, but we need not use much of these details here because our conclusions, as we shall see, depend little on them, but rather on the general form of representation (5)-(6) of a problem. In particular, the effective dependence of these unknown solutions on $\mathbf{g}_\pi$ is rather weak, for the significant terms of the sum over $\mathbf{g}_\pi$ in eq. (6c), which permits us not to specify this dependence (see also Appendix).

To obtain the complete solution to a problem, one should find the solutions of the modified Schrödinger equation, eq. (5), and then substitute them into the expression for $\psi_{\mathbf{g}_\pi}(\mathbf{r}_\sigma)$, after which the general solution can be written as

$$\Psi(\mathbf{r}) = \sum_n c_n \left[ \psi_{0n}(\mathbf{r}_\sigma) + \sum_{\mathbf{g}_\pi} \psi_{\mathbf{g}_\pi n}(\mathbf{r}_\sigma) \exp(i\mathbf{g}_\pi \mathbf{r}_\pi) \right] \exp(i\mathbf{K}_{\pi n} \mathbf{r}_\pi) =$$

$$= \sum_n c_n \exp(i\mathbf{K}_{\pi n} \mathbf{r}_\pi) \left[ 1 + \sum_{\mathbf{g}_\pi} \exp(i\mathbf{g}_\pi \mathbf{r}_\pi) \xi_{\mathbf{g}_\pi n}(\mathbf{r}_\sigma) \right] \psi_{0n}(\mathbf{r}_\sigma) , \quad (7)$$

where

$$\psi_{\mathbf{g}_\pi n}(\mathbf{r}_\sigma) = \xi_{\mathbf{g}_\pi n}(\mathbf{r}_\sigma) \psi_{0n}(\mathbf{r}_\sigma) \equiv \int_{s_\sigma} d\mathbf{r}_\sigma' x_{\mathbf{g}_\pi n}(\mathbf{r}_\sigma, \mathbf{r}_\sigma') \psi_0(\mathbf{r}_\sigma') ,$$

$$x_{\mathbf{g}_\pi n}(\mathbf{r}_\sigma, \mathbf{r}_\sigma') = V_{\mathbf{g}_\pi}(\mathbf{r}_\sigma') \sum_{n'} \frac{\psi^0_{\mathbf{g}_\pi n'}(\mathbf{r}_\sigma) \psi^{0*}_{\mathbf{g}_\pi n'}(\mathbf{r}_\sigma')}{\varepsilon_{\sigma n} - \varepsilon^0_{\mathbf{g}_\pi n'} - \varepsilon_{\pi g_\pi} - 2\cos\alpha_{\mathbf{g}_\pi}\sqrt{(E - \varepsilon_{\sigma n})\varepsilon_{\pi g_\pi}}} ,$$

(8)

the coefficients $c_n$ are to be determined from the boundary or initial conditions, as well as certain components of $\mathbf{K}$, $\varepsilon_{\mathbf{g}_\pi n}$ and $\mathbf{K}_{\pi n}$ are specified by eqs. (3) with $\varepsilon_\sigma = \varepsilon_{\sigma n}$, and $\{\psi_{0n}\}$, $\{\varepsilon_{\sigma n}\}$ are the complete sets of eigenfunctions and eigenvalues for the modified Schrödinger equation, eq. (5). And finally the main measured quantity, the probability density distribution (PDD), $\rho(\mathbf{r}) \equiv |\Psi(\mathbf{r})|^2$, can be expressed directly from eqs. (7), (8). The resulting general formula for it can be found elsewhere (see eq. (18) in ref. [17]).

Before analysing the results obtained, it would be not out of place to note that another practically important case, that of time-dependent periodical perturbation, $H_p = V_p(\mathbf{r}_\sigma, t)$, is effectively described by the same system of equations (2), where one should make the substitutions

$$\mathbf{r}_\pi \to t, \quad \mathbf{g}_\pi \to k \ (k \neq 0 \text{ is an integer}), \quad \varepsilon_{\sigma \mathbf{g}_\pi} \to \varepsilon_{\sigma k} \equiv \varepsilon_\sigma - \hbar\omega_\pi k,$$

the wave function being presented in the form

$$\Psi(\mathbf{r}_\sigma, t) = \exp(-i\varepsilon_\sigma t/\hbar) \left[ \psi_0(\mathbf{r}_\sigma) + \sum_k \psi_k(\mathbf{r}_\sigma) \exp(i\omega_\pi k t) \right] ,$$

and $\omega_\pi$ being the main frequency of the perturbation:

$$V_{\mathrm{p}}(\mathbf{r}_\sigma,t) = V_0(\mathbf{r}_\sigma) + \sum_k V_k(\mathbf{r}_\sigma)\exp(i\omega_\pi kt) \,.$$

This conclusion can be verified starting from the substitution of the total wave function above into the time-dependent Schrödinger equation. Then it is easily seen that this problem can also be reduced to solution of the modified Schrödinger equation (5), where the kernel of the effective potential can be expressed in a slightly different form,

$$V(\mathbf{r}_\sigma,\mathbf{r}_\sigma') = \sum_{k,n} \frac{V_{-k}(\mathbf{r}_\sigma)V_k(\mathbf{r}_\sigma')\psi^0_{kn}(\mathbf{r}_\sigma)\psi^{0*}_{kn}(\mathbf{r}_\sigma')}{\varepsilon_\sigma - \varepsilon^0_{kn} - \hbar\omega_\pi k} \,, \tag{6d}$$

and $\{\psi^0_{kn}(\mathbf{r}_\sigma)\}$, $\{\varepsilon^0_{kn}\}$ are determined from the auxiliary system of equations:

$$h_0(\mathbf{r}_\sigma)\psi_k(\mathbf{r}_\sigma) + \sum_{k'} V_{k-k'}(\mathbf{r}_\sigma)\psi_{k'}(\mathbf{r}_\sigma) = \varepsilon_{\sigma k}\psi_k(\mathbf{r}_\sigma) \,, \quad k,k' \neq 0. \tag{4b}$$

The general solution is (cf. eq. (7)):

$$\Psi(\mathbf{r}_\sigma,t) = \sum_n c_n \exp(-i\varepsilon_{\sigma n}t/\hbar)\left[1 + \sum_k \exp(i\omega_\pi kt)\xi_{kn}(\mathbf{r}_\sigma)\right]\psi_{0n}(\mathbf{r}_\sigma) \,,$$

with the integral kernel $x_{kn}(\mathbf{r}_\sigma,\mathbf{r}_\sigma')$ of the operators $\xi_{kn}(\mathbf{r}_\sigma)$ given by

$$x_{kn}(\mathbf{r}_\sigma,\mathbf{r}_\sigma') = V_k(\mathbf{r}_\sigma') \sum_{n'} \frac{\psi^0_{kn'}(\mathbf{r}_\sigma)\psi^{0*}_{kn'}(\mathbf{r}_\sigma')}{\varepsilon_{\sigma n} - \varepsilon^0_{kn'} - \hbar\omega_\pi k} \,,$$

and $\{\varepsilon_{\sigma n}\}$, $\{\psi_{0n}(\mathbf{r}_\sigma)\}$ determined from eq. (5). Starting from these formulas one obtains any desired measurable quantity, similar to the time-independent formalism.

It is clear that the time-dependent case is a generalisation of the well-known group of the model "kicked" systems with similar behaviour described by the basic standard map [20] (called also the standard model, see e. g. [2,3]). This model corresponds to the δ-like periodic kicks (i. e. the components $V_k(\mathbf{r}_\sigma) = V_0(\mathbf{r}_\sigma)$ do not depend on $k$) and also to some particular choices of the potential dependence on $\mathbf{r}_\sigma \equiv x$ (typically $V_0(x)$, $V_k(x) \propto -\cos(x)$). The proposed generalisation can serve thus as a more realistic representation for many particular physical systems studied with the standard model. By analogy one may designate our time-dependent case as the generalised kicked oscillator. Because of its similarity to the time-independent case we now continue the analysis of the latter mentioning the differences between the two where necessary.

## 2.2. Fundamental dynamic multivaluedness

It is not surprising to see, from the above expressions, that the EP method thus formulated cannot provide directly the exact solutions and is nothing but another formulation of a problem. However, we can show now that for the chaotic systems it is this representation that is much more relevant than the ordinary one (in this case, the Schrödinger equation with the ordinary potential, $V(\mathbf{r})$). It permits one to obtain, in a natural and self-consistent manner, a basic source of randomness and complexity in such systems and then to study, in relation to these fundamental concepts, their particular chaotic properties in terms of observable quantities.

We see from the above formulas that the experimentally measured PDD is determined by the dynamics in EP. The distinctive property of the latter, and the most important one as far as the dynamical chaos is concerned, is its self-consistent dependence on the energy eigenvalues to be determined. This dependence appears in the explicit form when one tries to find the energy eigenvalues $\{\varepsilon_{\sigma n}\}$ from the modified Schrödinger equation, eq. (5), while using the expressions (6c,d) for the integral kernel of $V_{\text{eff}}(\mathbf{r}_\sigma)$ which depends itself on $\varepsilon_\sigma$. Note that this property is restricted solely to the full non-perturbative EP formalism studied here as opposed to its various perturbative versions eventually used in many applications [16] including the problem of quantum chaos [21,22] (see also section 5.III). In fact, we deal here with the intrinsic effective nonlinearity of a chaotic system which is not taken into account either by the ordinary, non-modified formalism, or by the perturbative approaches (further discussion of the effective nonlinearity can be found in sections 5.III, 6.III, and 9.2.IV). Now we are going to show that this peculiar property leads to the conclusion that, instead of $N_\sigma$ eigenvalues and eigenfunctions for the Schrödinger equation with the "non-chaotic" potential $V_0(\mathbf{r}_\sigma)$, one obtains up to

$$N_{\max} = (N_\pi N_\pi' + 1)N_\sigma \qquad (9a)$$

solutions for the equation with EP, where $N_\pi$ and $N_\pi'$ are the numbers of terms in the sums over $\mathbf{g}_\pi$ and $n$, respectively, in eq. (6c). Among these $N_{\max}$ solutions, $N_0 = (N_\pi' + 1)N_\sigma$ solutions correspond to the normal set of eigenfunctions for the full-dimensional Schrödinger equation with the potential $V(\mathbf{r})$ (in a standard situation there should be $N_\pi = N_\pi'$, neglecting the $\varepsilon^0_{\mathbf{g}_\pi n}$ dependence on $\mathbf{g}_\pi$, see Appendix). The *additional* growth of the number of solutions by

$$N_\Delta \equiv N_{\max} - N_0 = N_\pi'(N_\pi - 1)N_\sigma \geq N_\sigma$$

cannot be explained in terms of ordinary splitting effects. These extra solutions could be, in principle, spurious, unphysical ones (e. g. unstable). This can indeed happen for *some* ranges of parameters, and this is precisely one of the mechanisms of regularity in chaotic quantum systems, studied below. However, it is difficult to imagine, and we confirm these doubts below, that this may be the case for all parameter values. We shall see that most often at least some of these additional solutions are quite real and observable.

Before proving these statements, note that it will be convenient to count the solutions in terms of $N_\sigma$: $N_r \equiv N/N_\sigma$, where $N$ is the total number of eigenvalues (eigenfunctions) in a general case;

$$N_r^{\max} \equiv N_{\max}/N_\sigma = N_\pi N_\pi' + 1; \quad N_r^0 \equiv N_0/N_\sigma = N_\pi' + 1; \tag{9b}$$

$N_r^\Delta \equiv N_r^{\max} - N_r^0 = N_\Delta/N_\sigma = N_\pi'(N_\pi - 1) \geq 1$. One may introduce also $n_r \equiv N_r/N_r^0$ and

$$n_r^{\max} \equiv N_r^{\max}/N_r^0 = (N_\pi N_\pi' + 1)/(N_\pi' + 1) \to N_\pi \text{ for } N_\pi' \gg 1 . \tag{9c}$$

The statements above, concerning the number of solutions, can be verified by at least three different ways giving all the same result, eqs. (9): one can count solutions directly by analysing the system of equations corresponding to eq. (5) in the momentum or other suitable representation; one can study a problem by a graphical method; at last, reasonable and easily treated approximations of eqs. (5)-(6) can be proposed. Consider now, in consecutive order, these arguments in detail.

The momentum representation of eq. (5) is

$$h_{\mathbf{g}_\sigma}\psi_{\mathbf{g}_\sigma} + \sum_{\mathbf{g}_\sigma'} V^{\text{eff}}_{\mathbf{g}_\sigma,\mathbf{g}_\sigma'}\psi_{\mathbf{g}_\sigma'} = \psi_{\mathbf{g}_\sigma}\varepsilon_\sigma , \tag{10}$$

where, for the case of $V(\mathbf{r})$ periodic in $\mathbf{r}_\sigma$, the notations $h_{\mathbf{g}_\sigma}$, $\psi_{\mathbf{g}_\sigma}$, and $V^{\text{eff}}_{\mathbf{g}_\sigma,\mathbf{g}_\sigma'}$ hold respectively for the Fourier representations of the free-motion Hamiltonian, the periodic part of the wave function $\psi_0(\mathbf{r}_\sigma)$ and the EP kernel $V(\mathbf{r}_\sigma,\mathbf{r}_\sigma')$,

$$V^{\text{eff}}_{\mathbf{g}_\sigma,\mathbf{g}_\sigma'} = \sum_{\mathbf{g}_\pi,n} \frac{\sum_{\mathbf{g}_\sigma'',\mathbf{g}_\sigma'''} V_{-\mathbf{g}_\pi,\mathbf{g}_\sigma-\mathbf{g}_\sigma''} V_{\mathbf{g}_\pi,\mathbf{g}_\sigma''+\mathbf{g}_\sigma'''} \Psi_{\mathbf{g}_\pi\mathbf{g}_\sigma''} \Psi^*_{\mathbf{g}_\pi\mathbf{g}_\sigma'''}}{\varepsilon_\sigma - \varepsilon_{\mathbf{g}_\pi n} - \varepsilon_{\pi\mathbf{g}_\pi} - 2\cos\alpha_{\mathbf{g}_\pi}\sqrt{(E-\varepsilon_\sigma)\varepsilon_{\pi\mathbf{g}_\pi}}} , \tag{11}$$

while the (straightforward) details of the definitions of the participating Fourier-components of the total potential and wave function are not important in the context. For the case of $V(\mathbf{r})$ non periodic in $\mathbf{r}_\sigma$ one can use expansion in any appropriate complete set of orthogonal functions (for example, the set of eigenfunctions for $H_0$), which gives effectively the same result with $\mathbf{g}_\sigma$, $\mathbf{g}_\sigma'$, ... in eqs. (10),(11) enumerating these functions. Now if we neglect $\varepsilon_\sigma$ under the root in the denominator of eq. (11) for larger $E$, $E \gg \varepsilon_\sigma$, then diagonalisation of system (10) is evidently reduced to solution of an algebraic equation in $\varepsilon_\sigma$ of degree $N_{\max}$ defined by eq. (9a). The latter statement is thus valid for the *maximum attainable value* of the master equation degree and the number of solutions of the modified Schrödinger equation, while the influence of the neglected $\varepsilon_\sigma$ term is considered in detail below, in section 3.II.

In what follows we call this *additional* splitting of solutions and, correspondingly of the effective Hamiltonian, potential or other relevant dynamical function, the fundamental multivaluedness of dynamical functions (FMDF). It includes the ordinary dimensional splitting into the *complete* set of $N_r^0$ solutions as *one* component, but also, possibly, other similar complete components. We shall call the *i*-th component the *i*-th *realisation* of a dynamical system (or of a problem) and designate it by $\Re_i$; it effectively includes, besides the corresponding complete set of solutions, also the corresponding components of EP and PDD:

$$\Re_i \equiv \{\{\varepsilon_{\sigma n}\}^i, \{\psi_{0n}(\mathbf{r}_\sigma)\}^i, V_{\text{eff}}^i(\mathbf{r}_\sigma), \rho_i(\mathbf{r})\} \quad (i = 1,2,...,N_\Re) \;, \tag{12}$$

where $N_\Re \geq 1$ is the total number of realisations in their *ensemble* thus obtained. It is formally limited from above:

$$N_\Re \leq N_\Re^{\max} \leq N_r - N_r^0 + 1 \leq N_r^{\max} - N_r^0 + 1 = N_\pi'(N_\pi - 1) + 1$$

(we shall see below that actually $N_\Re = N_\pi$).

Now we can present another, the most convincing and meaningful, evidence supporting the reality of the fundamental multivaluedness of solutions and based on the graphical analysis of eq. (5). We first rewrite eq. (5) for certain *n*-th eigenvalue:

$$[h_0(\mathbf{r}_\sigma) + V_0(\mathbf{r}_\sigma)]\psi_{0n}(\mathbf{r}_\sigma) + \int_{s_\sigma} d\mathbf{r}_\sigma' V_n(\mathbf{r}_\sigma,\mathbf{r}_\sigma')\psi_{0n}(\mathbf{r}_\sigma') = \varepsilon_{\sigma n}\psi_{0n}(\mathbf{r}_\sigma) \;.$$

Multiplying it by $\psi_{0n}^*(\mathbf{r}_\sigma)$ and integrating over $\mathbf{r}_\sigma$ we arrive at the following formulation of a problem:

$$V_{nn}(\varepsilon_{\sigma n}) = \varepsilon_{\sigma n} - \overset{0}{\varepsilon}_{\sigma n} \;, \tag{13}$$

where

$$V_{nn}(\varepsilon_{\sigma n}) \equiv \sum_{\mathbf{g}_\pi, n'} \frac{|V_{\mathbf{g}_\pi}^{nn'}|^2}{\varepsilon_{\sigma n} - \overset{0}{\varepsilon}_{\mathbf{g}_\pi n'} - \varepsilon_{\pi g_\pi} - 2\cos\alpha_{\mathbf{g}_\pi}\sqrt{(E - \varepsilon_{\sigma n})\varepsilon_{\pi g_\pi}}}, \tag{14a}$$

$$V_{\mathbf{g}_\pi}^{nn'} \equiv \int_{s_\sigma} d\mathbf{r}_\sigma \psi_{\mathbf{g}_\pi n'}^{0*}(\mathbf{r}_\sigma) V_{\mathbf{g}_\pi}(\mathbf{r}_\sigma) \psi_{0n}(\mathbf{r}_\sigma) \;,$$

and

$$\overset{0}{\varepsilon}_{\sigma n} \equiv \int_{s_\sigma} d\mathbf{r}_\sigma \psi_{0n}^*(\mathbf{r}_\sigma)[h_0(\mathbf{r}_\sigma) + V_0(\mathbf{r}_\sigma)]\psi_{0n}(\mathbf{r}_\sigma) \;.$$

Function $V_{nn}(\varepsilon_{\sigma n})$ possesses a number of singularities determined by the zeros of the denominators of each term of the double sum in eq. (14a). The zero, and thus the corresponding singularity position, for a term with certain $n'$ and $\mathbf{g}_\pi$ is easily found to be at $\varepsilon_{\sigma n} = \varepsilon^{\pm}_{\sigma n'}(\mathbf{g}_\pi)$, where

$$\varepsilon^{\pm}_{\sigma n'}(\mathbf{g}_\pi) = \varepsilon_{\mathbf{g}_\pi n'} + \varepsilon_{\pi g_\pi}(1 - 2\cos^2\alpha_{\mathbf{g}_\pi}) + 2\cos\alpha_{\mathbf{g}_\pi}\sqrt{\varepsilon_{\pi g_\pi}(E - \varepsilon_{\pi g_\pi}\sin^2\alpha_{\mathbf{g}_\pi} - \varepsilon_{\mathbf{g}_\pi n'})}. \tag{15a}$$

The superscript "±" in the notations above serves to remind us about a feature important for the following analysis: for each $g_\pi$ there are two terms in the sum (14a) with the opposite directions of the vector $\mathbf{g}_\pi$ corresponding to change of sign of $\cos\alpha_{\mathbf{g}_\pi}$ in eq. (15a). This can be expressed in a straightforward fashion for the simple and rather common case of one-dimensional perturbation ($\mathbf{r}_\pi \equiv z$):

$$\varepsilon^{\pm}_{\sigma n'}(\pm|g_\pi|) = \varepsilon_{g_\pi n'} - \varepsilon_{\pi g_\pi} \pm 2\sqrt{\varepsilon_{\pi g_\pi}(E - \varepsilon_{g_\pi n'})}.$$

For the case of time-dependent perturbation (the generalised kicked oscillator) one obtains, using the substitution procedure described above, the same eq. (13) with

$$V_{nn}(\varepsilon_{\sigma n}) \equiv \sum_{k,n'} \frac{|V_k^{nn'}|^2}{\varepsilon_{\sigma n} - \varepsilon^0_{kn'} - \hbar\omega_\pi k}, \tag{14b}$$

where $V_k^{nn'} \equiv V_{\mathbf{g}_\pi}^{nn'}\big|_{\mathbf{g}_\pi = k}$ and the positions of singularities are determined as

$$\varepsilon^{\pm}_{\sigma n'}(k) = \varepsilon_{kn'} + \hbar\omega_\pi k = \varepsilon_{kn'} \pm \hbar\omega_\pi|k|. \tag{15b}$$

Now to apply our graphical analysis we plot, in Fig. 1 (p. 15), the left- and right-hand sides of eq. (13) vs $\varepsilon_{\sigma n}$ taken as continuous independent variable. The solutions are found then as abscissae of the points of intersection of the curves corresponding to the two functions plotted. As is seen from the figure, representing eq. (13) for two characteristic cases of parameter values (see section 3.II), function $V_{nn}(\varepsilon_{\sigma n})$ consists of many branches due to the sums over $n'$ and $\mathbf{g}_\pi$ of divergent terms. The points of divergence are designated by the corresponding vertical asymptotes, their positions on the horizontal axis being determined by eqs. (15). It is this plurality of branches which gives multiple solutions for $\varepsilon_{\sigma n}$ (we enumerate them by another index, $j$: $\varepsilon^j_{\sigma n}, j > 0$) instead of the single one, $\varepsilon_{\sigma n} = \varepsilon^0_{\sigma n}$, for the Schrödinger equation with the 'regular' potential $V_0(\mathbf{r}_\sigma)$. One part of this splitting can be explained by the ordinary multiplication of the number of eigenvalues and eigenfunctions due to the addition of the degrees of freedom corresponding to $\mathbf{r}_\pi$ to those of $\mathbf{r}_\sigma$. However, there is also another part of the splitting which just represents the FMDF described above. Thus for the case of Fig. 1(a), where we have restricted ourselves only to two terms of each of the sums over $n'$ and $\pm|g_\pi|$ (i. e. $N_\pi = 4$, $N_\pi' = 2$), in the absence of this additional splitting one could not obtain more than $N_r^0 = N_\pi' + 1 = 3$ solutions, whereas actually one can count $N_r = N_r^{\max} = 9$ solutions for $\varepsilon_{\sigma n}$ in full agreement with the general expression, eq. (9b).[*)] To deduce eq. (9) from our graphical analysis, we just note that, as follows from eqs. (14), (15), the number of asymptotes is equal to $N_\pi N_\pi'$, and then, as is clear from the figure, the number of points of intersection of the two curves, and thus of the solutions, is $N_r^{\max} = N_\pi N_\pi' + 1$. We shall see in section 3.III why and how this *maximum* number of solutions can diminish down to $N_r^0$ giving rise to qualitative changes in the chaotic behaviour of a system. Figure 1(b) reproduces these results for another characteristic parameter values and for the value $N_\pi' = 3$ (we continue this analysis in section 3.III). It can be especially clearly seen from configuration in this figure that the eigen-solutions are grouped so as to form $N_\Re = N_\pi$ realisations. Note that the constant $\varepsilon_{\sigma n}^0$ adjusts itself, in a self-consistent manner, to its particular value for each realisation.[#)]

It would be useful to introduce the notion of distance between two realisations. For our purposes it is quite sufficient to define it approximately as the average, or typical, magnitude of the difference between the values of the corresponding branches of EP. Then it follows from the above analysis that the separation between realisations (i. e. the distance between the neighbouring ones) is generally between the energy-level separation for the unperturbed potential, $\Delta \varepsilon_\sigma$, at maximum (in the situation of global regularity, Fig. 1(b), see section 3.III), and the minimum of $2(\varepsilon_{\pi g_{\pi 0}} E)^{1/2} = 2\pi \hbar \sqrt{2E/m}/d_\pi$ (in the situation of global chaos, Fig. 1(a)). We see that the finite separation between the realisations, proportional to $\hbar$, (i. e. their *discreteness*) is a quantum effect. As will be shown in the next section, it has the important physical consequences. It means also that in the semiclassical situation one has, at the scale of the characteristic potential values, many closely separated realisations forming a quasi-continuous distribution. Of course, these generic rules do not exclude the existence of particular realisation separations of greater (or smaller) magnitudes.

---

[*)] Note that for convenience of illustration we consider cases with $N_\pi \neq N_\pi'$, whereas normally one should have $N_\pi = N_\pi'$; however, this choice does not influence the obtained conclusions.

[#)] As the value of $\varepsilon_{\sigma n}^0$ may vary for different solutions $\varepsilon_{\sigma n}^j$, the segments of the line $\varepsilon_{\sigma n} - \varepsilon_{\sigma n}^0$, intersecting the respective branches of the function $V_{nn}(\varepsilon_{\sigma n})$ in Fig. 1, can be slightly displaced vertically one relative to another. However, this will not produce any significant changes on the scale of our schematical representation, neither in the conclusions obtained, and we do not show these secondary details to avoid unnecessary complication.

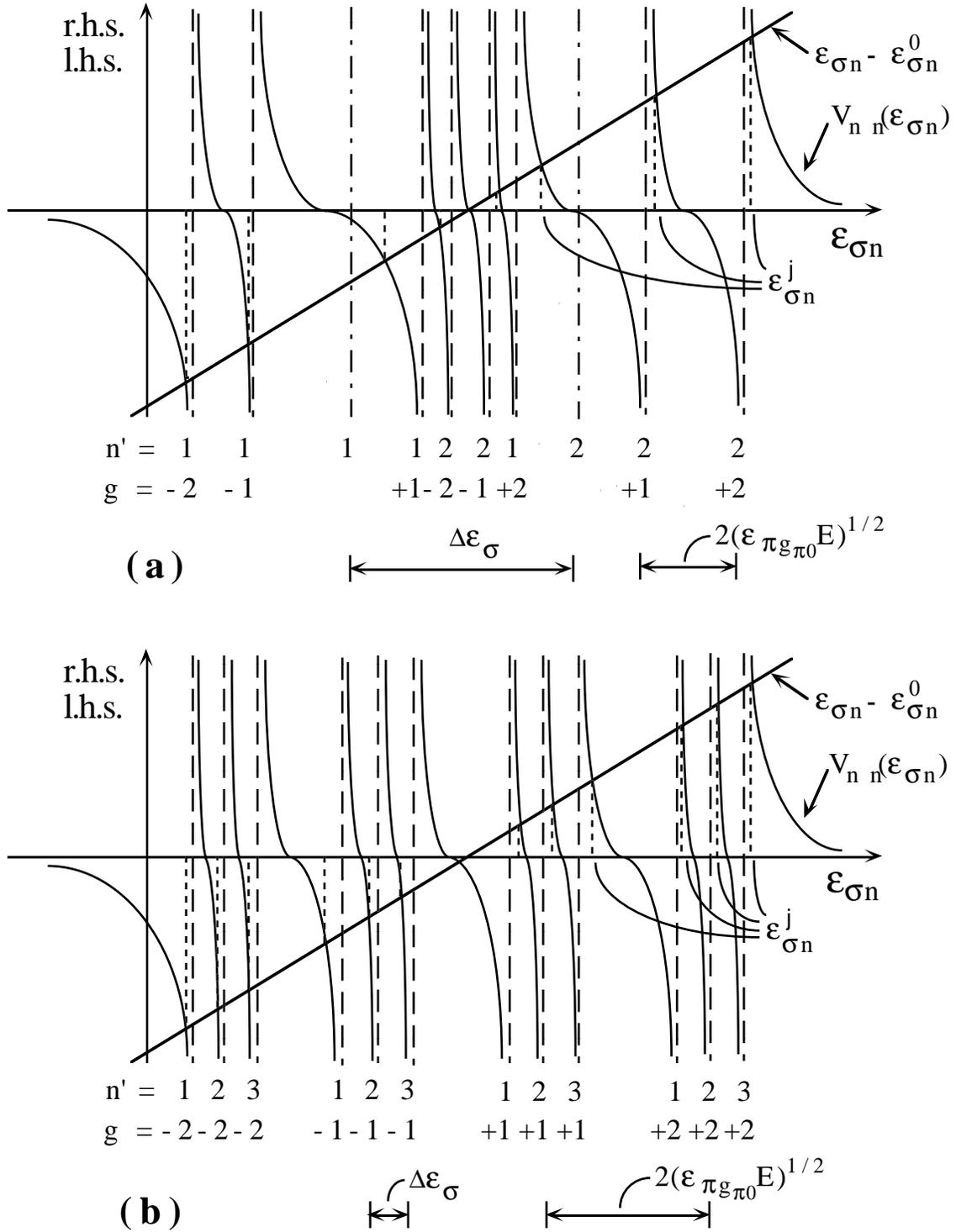

**Fig. 1**. Graphical solution of the modified Schrödinger equation, eq. (13), for the Hamiltonian system with periodic perturbation in the parameter domains of global chaos **(a)**, and global regularity **(b)**. We plot the left- and the right-hand sides of eq. (13) vs eigenvalue to be determined, $\varepsilon_{\sigma n}$. The illustration corresponds to the case $E \gg \varepsilon_{\sigma n}$ (see text) and the following numbers of terms in the sums over $g_\pi$ and $n'$ in the expression for $V_{nn}(\varepsilon_{\sigma n})$, eq. (14a): $N_\pi = 4$, $N_\pi' = 2$ (a), and $N_\pi = 4$, $N_\pi' = 3$ (b). The asymptotes of the function $V_{nn}(\varepsilon_{\sigma n})$ are shown by the dashed lines with the respective values of $g$ and $n'$ marked at the lower end of each of them, where the non-zero integer $g$ enumerates $g_\pi$: $g_\pi = 2\pi g/d_\pi$ ($g_{\pi 0} = 2\pi/d_\pi$). Two vertical dash-dotted lines in (a) correspond to the values $\varepsilon_{\sigma n} = \varepsilon^0_{g_\pi n'}$.

Finally, we refer to our paper [17] for the detailed description of a simple model for eq. (13) using another form of $V_{nn}(\varepsilon_{\sigma n})$. This model corresponds to the simplest choice $N_\pi = 2$, $N_\pi' = 1$, $N_\sigma = 1$ and is reduced to a cubic equation that can be directly analysed including the dependence on parameters. Despite apparently rough approximations, the model is shown to be qualitatively reasonable for a large range of practical cases. We do not reproduce it here because effectively it represents a particular case of a more general analysis performed above. Nonetheless, it gives a transparent illustration of the appearance of extra solutions and therefore confirms once more the above general analysis results.

# Appendix

The formalism of the fundamental multivaluedness, providing true quantum chaos in Hamiltonian systems, is based on the division of an "unresolvable" problem into two parts. One of them, represented by the modified Schrödinger equation, eq. (5), is relatively simply structured, easily tractable and gives finally the fundamental dynamic uncertainty in the form of problem splitting into plural realisations. The second part is represented by the auxiliary system of equations, eqs. (4), which is more difficult for the detailed analysis (for example, because of the large number of the participating equations), but influences the first part and the solution of a problem in a less direct, largely qualitative way. It provides a necessary 'material support' for the fundamental uncertainty and plays a minor (though irreducible) role for the basic origins of dynamical randomness. Nonetheless, one needs to know the general properties of the auxiliary system solutions as well as the effective approximations that could be used in practical applications of the formalism of FMDF. We demonstrate these features of the auxiliary system of equations for the more transparent time-dependent case, while the time-independent problem can be analysed in the same way and with the same results accompanied by minor technical changes.

A useful approximation for the auxiliary system, eq. (4b), can be obtained if we rewrite it in the form:

$$h_0(\mathbf{r}_\sigma)\psi_k^0(\mathbf{r}_\sigma) + V_0(\mathbf{r}_\sigma)\psi_k^0(\mathbf{r}_\sigma) = \varepsilon_k^0 \psi_k^0(\mathbf{r}_\sigma) - \sum_{k' \neq k} V_{k'}(\mathbf{r}_\sigma)\psi_{k-k'}^0(\mathbf{r}_\sigma) \qquad (A.1)$$

(we recall that by definition $k, k' \neq 0$). We can apply now the same EP formalism that was used in section 2.1 to obtain the modified Schrödinger equation, regarding the sum in the right-hand side of eq. (A.1) as a perturbation. The result differs from that of eq. (5) in two points: the EP in this case evidently depends on $k$, and it cannot be obtained in a closed form, but only as a series.[*] Omitting the detailed calculations quite similar to those of section 2.1, we write down the first term of this series for the EP kernel:

$$V_k(\mathbf{r}_\sigma, \mathbf{r}_\sigma') = \sum_n \sum_{k' \neq k} \frac{V_{k'}(\mathbf{r}_\sigma) V_{-k'}(\mathbf{r}_\sigma') \psi_{0n}^0(\mathbf{r}_\sigma) \psi_{0n}^{0*}(\mathbf{r}_\sigma')}{\varepsilon_{k-k'}^0 - \varepsilon_{0n}^0 + \hbar\omega_\pi(k - k')}, \qquad (A.2)$$

---

[*] The latter feature seems to have a fundamental implication involved with the fractal structure of dynamic complexity. The dependence of the higher-order terms on the eigenvalue to be found accounts for the appearance of the respective higher-order fractal branches, together with similar dependence of another quantity, the 'averaged energy addition' (see section 4.III for more details). However, these details should not directly influence the structure of the *current* level of the structure of chaos.

where $\{\varepsilon^0_{0n}\}$ and $\{\psi^0_{0n}\}$ are the eigenvalues and the eigenfunctions for the unperturbed potential:

$$h_0(\mathbf{r}_\sigma)\psi^0_{0n}(\mathbf{r}_\sigma) + V_0(\mathbf{r}_\sigma)\psi^0_{0n}(\mathbf{r}_\sigma) = \varepsilon^0_{0n}\psi^0_{0n}(\mathbf{r}_\sigma) \ . \tag{A.3}$$

As $\varepsilon^0_k = \varepsilon^0_\sigma - \hbar\omega_\pi k$, one can rewrite eq. (A.2) in the form:

$$V_k(\mathbf{r}_\sigma,\mathbf{r}_\sigma') = \sum_n \frac{\psi^0_{0n}(\mathbf{r}_\sigma)\psi^{0*}_{0n}(\mathbf{r}_\sigma')\sum_{k'\neq k}V_{k'}(\mathbf{r}_\sigma)V_{-k'}(\mathbf{r}_\sigma')}{\varepsilon^0_k - \varepsilon^0_{0n} + \hbar\omega_\pi k} \ . \tag{A.4}$$

This leads to the following graphical-analysis representation (see eqs. (13), (14)) of the corresponding modified Schrödinger equations:

$$V_{nn}(k,\varepsilon^0_{kn}) = \varepsilon^0_{kn} - \varepsilon^{00}_{kn} \ , \tag{A.5}$$

where

$$V_{nn}(k,\varepsilon^0_{kn}) \equiv \sum_{n'} \frac{\sum_{k'\neq k}|V^{nn'}_{k'}|^2}{\varepsilon^0_{kn} - \varepsilon^0_{0n'} + \hbar\omega_\pi k} \ , \tag{A.6}$$

$$V^{nn'}_k \equiv \int_{s_\sigma} d\mathbf{r}_\sigma \psi^{0*}_{0n'}(\mathbf{r}_\sigma) V_k(\mathbf{r}_\sigma)\psi^0_{kn}(\mathbf{r}_\sigma) \ ,$$

and

$$\varepsilon^{00}_{kn} \equiv \int_{s_\sigma} d\mathbf{r}_\sigma \psi^{0*}_{kn}(\mathbf{r}_\sigma)[h_0(\mathbf{r}_\sigma) + V_0(\mathbf{r}_\sigma)]\psi^0_{kn}(\mathbf{r}_\sigma) \ .$$

The singular denominators in eq. (A.6) determine the asymptote positions (section 2.2), and we see that, contrary to the case of eqs. (13), (14), now there is no additional splitting of solutions: one has $N_\sigma$ asymptotes for each $k$ which provide only one realisation (at *this* level of complexity, cf. section 4.III). Moreover, the obtained eigenvalue set depends little on $k$ (it is rather shifted as a whole with changing $k$). The weak dependence on $k$ of the corresponding EP can be also traced from eq. (A.4). If one adds another terms of the series for the EP, this will not change our qualitative results because each of these terms contains the same denominator as eq. (A.6), but raised to the corresponding power, and the respective additional summations of terms quadratic in $V_k(\mathbf{r}_\sigma)$ in the numerator. Note, by the way, that this situation gives an idea about how one can reproduce regularity, within our method, for an arbitrary *integrable* system.

Another approximation and a way of analysis of the auxiliary system can be obtained if we add the same quantity, $\psi_k^0(\mathbf{r}_\sigma) \sum_{k' \neq k} V_{k'}(\mathbf{r}_\sigma)$, to both sides of the initial eq. (A.1):

$$h_0(\mathbf{r}_\sigma)\psi_k^0(\mathbf{r}_\sigma) + \left[V_0(\mathbf{r}_\sigma) + \sum_{k' \neq k} V_{k'}(\mathbf{r}_\sigma)\right]\psi_k^0(\mathbf{r}_\sigma) =$$
$$= \varepsilon_k^0 \psi_k^0(\mathbf{r}_\sigma) - \sum_{k' \neq k} V_{k'}(\mathbf{r}_\sigma)\left[\psi_{k-k'}^0(\mathbf{r}_\sigma) - \psi_k^0(\mathbf{r}_\sigma)\right] . \quad (A.7)$$

If we suppose that the effective potential,

$$V_1(\mathbf{r}_\sigma) \equiv V_0(\mathbf{r}_\sigma) + \sum_{k' \neq k} V_{k'}(\mathbf{r}_\sigma) = V(\mathbf{r}_\sigma, t=0) - V_k(\mathbf{r}_\sigma) , \quad (A.8)$$

as well as the solution, $\psi_k^0(\mathbf{r}_\sigma)$, depend only weakly on $k$, this will provide a self-consistent approximation for eq. (A.7):

$$h_0(\mathbf{r}_\sigma)\psi_k^0(\mathbf{r}_\sigma) + V_1(\mathbf{r}_\sigma)\psi_k^0(\mathbf{r}_\sigma) = \varepsilon_k^0 \psi_k^0(\mathbf{r}_\sigma) . \quad (A.9)$$

This approximation is evidently more precise for the $\delta$-like perturbation dependence on $t$ containing many sufficiently large higher harmonics (in the limit it coincides with the $\delta$-functional kicks of the standard model). Contrary to this, the previous series expansion of the EP is more relevant to the case of quasi-harmonic perturbation. However, the qualitative properties of the auxiliary system solutions are similar for both approximations. The difference is determined by the effective potential which changes approximately between $V_0(\mathbf{r}_\sigma)$ for a harmonic perturbation and $V_1(\mathbf{r}_\sigma)$ for $\delta$-like kicks. These two limiting EP models show weak dependence on k and differ mostly by their amplitudes: $\Delta V_1 \gg \Delta V_0$. In this sense one may say that the EP follows, to some degree, the form of the perturbation.

In the time-independent case similar analysis leads to analogous results concerning the qualitative properties of the auxiliary system solutions. We do not reproduce it here noting only that the detailed expressions deal with $K_\pi$ rather than $\varepsilon_\sigma$ (see eqs. (3)).

# References

[1]  E. Ott, *Chaos in dynamical systems* (Cambridge Univ. Press, Cambridge, 1993).

[2]  A.J. Lichtenbegr and M.A. Lieberman, *Regular and Stochastic Motion* (Springer-Verlag, New-York, 1983).

[3]  G.M. Zaslavsky, *Chaos in dynamical systems* (Harwood Academic Publishers, London, 1985).

[4]  M.C. Gutzwiller, *Chaos in Classical and Quantum Mechanics* (Springer-Verlag, New York, 1990).

[5]  B. Eckhardt, Phys. Rep. **163**, 205 (1988).

[6]  *Chaos and Quantum Physics*, edited by M.J. Giannoni, A. Voros, and J. Zinn-Justin (North-Holland, Amsterdam, 1991).

[7]  F. Haake, *Quantum signatures of chaos* (Springer, Berlin, 1991).

[8]  *Quantum and Chaos: How Incompatible ?* Proceedings of the 5th Yukawa International Seminar (Kyoto, August 1993), Ed. K. Ikeda, Progress Theor. Phys. Suppl. No. 116 (1994); W.-M. Zhang and D. H. Feng, Phys. Rep. **252**, 1 (1995).

[9]  E.B. Bogomolny, Nonlinearity **5**, 805 (1992).

[10]  O. Bohigas, in *Chaos and Quantum Physics* (ref. [6]); O. Bohigas, S. Tomsovic, and D. Ullmo, Phys. Rep. **223** (2), 45 (1993).

[11]  J. Ford and G. Mantica, Am. J. Phys. **60**, 1086 (1992); see also J. Ford, G. Mantica, and G. H. Ristow, Physica D **50**, 493 (1991); J. Ford and M. Ilg, Phys. Rev. A **45**, 6165 (1992); J. Ford, in *Directions in Chaos*, edited by Hao Bai-lin (World Scientific, Singapore, 1987), Vol. 1, p. 1; J. Ford, ibid., p. 129.

[12]  M.V. Berry, in *Chaos and Quantum Physics* (ref. [6]).

[13]  B.V. Chirikov, in *Chaos and Quantum Physics* (ref. [6]).

[14]  *Quantum Chaos - Quantum Measurement* (Kluwer, Dortrecht, 1992).

[15]  I. Prigogine, Physics Reports **219**, 93 (1992).

[16]  P.H. Dederichs, Solid State Phys. **27**, 136 (1972).

[17]  A.P. Kirilyuk, Nuclear Instrum. Meth. **B69**, 200 (1992).

[18]  A.I. Akhiezer, V.I. Truten', and N.F. Shul'ga, Phys. Rep. **203**, 289 (1991).

[19]  G. Casati, I. Guarneri, and D. Shepelyansky, Physica A **163**, 205 (1990).

[20]  B.V. Chirikov, Phys. Rep. **52**, 263 (1979); see also ref. [13].

[21]  G. Hose and H.S. Taylor, J. Chem. Phys. **76**, 5356 (1982).

[22]  G. Hose, H.S. Taylor, and A. Tip, J. Phys. A **17**, 1203 (1984).